\documentclass{elsart}
\usepackage{graphicx}

\usepackage{amssymb}

\begin{document}

\begin{frontmatter}

\title{Hot-border effects, ordering, and avalanches
in planar arrays of superheated superconducting granules
}

\author{A. Pe\~{n}aranda}
\ead{angelina@fa.upc.es},
\author{L. Ram\'{\i}rez-Piscina}

\address{
Departament de F\'{\i}sica Aplicada,\\
Universitat Polit\`{e}cnica de Catalunya,\\
Doctor Mara\~{n}\'on 44, E-08028 Barcelona, SPAIN.}

\begin{abstract}
We present results of simulations of transitions in planar arrays
of superheated superconducting granules  (PASS) immersed in an
external magnetic field parallel to the system. Analysis of the
behaviour of the system when the external field is slowly
increased from zero shows the existence of an interval of external
fields for which no transitions are produced. This gap is found to
be a consequence of ordering induced by transitions. However, this
ordering is different for diluted and for concentrated systems.
Avalanches of different sizes with distributions depending on
concentration are observed.

\end{abstract}

\begin{keyword}
A. Superconductors. \sep D. Phase transitions.

\PACS 41.20.G \sep 29.40.Ym \sep 74.80.Bj
\end{keyword}
\end{frontmatter}

\section{Introduction}

Experiments on dispersions of superheated superconducting granules
(SSG) have investigated the feasibility of these systems as
detectors of neutrinos, dark matter and other particles
\cite{girard0}. In these devices an ensemble of superconducting
type I microgranules are maintained in a metastable superheated
state by adequate conditions of temperature and external field. If
a particle or radiation with enough energy collides with a
microgranule it can produce a transition to the normal conducting
state. The disappearance of the Meissner effect, inherent to the
transition, produces a change in the magnetic flux that can be
detected by a magnetometer.

Despite their promising applications, some intrinsic features of SSG limit
the efficiency of these detectors. The metastability of each grain of the
suspension depends on its shape and surface quality and on the diamagnetic
interactions with the other microgranules. Dispersion in size combined with
these effects produces a spread of transition fields of about $20\%$, which
affects the accuracy of these devices.

Planar systems of microgranules arrayed in a regular configuration (PASS)
have been presented as an alternative to SSG  to reduce this
spread \cite{PASS}. The present manufacturing techniques of PASS devices
can only produce arrays of a relatively small size, but with good
sensitivity in both energy and position. The regular arrangement allows
the control of magnetic interactions. Furthermore, the photolithographic
procedure employed in their manufacture enables microgranules with
excellent uniformity in shape and size to be obtained.
Although these characteristics
do not completely eliminate the dispersion of superheated transition field
values,
reductions of an order of magnitude have been observed \cite{PASS}.
$ \alpha $
On the other hand, the regular spatial configuration and the
greater uniformity in shape and size of these systems suggest the
possibility of the presence of avalanches that could amplify the
magnetic signal. Avalanches are a consequence of diamagnetic
interactions between grains, which screen the magnetic field of
neighbouring granules. The transition of a simple grain can reduce
or eliminate this screening  in such a way that the maximum
surface field of the nearest granules can increase. Consequently,
transitions in several neighbours can occur almost simultaneously.
Diffusion of released heat, inherent to any first-order transition
could also produce transitions in neighbouring grains. This
avalanche effect could be useful for devices with very small
granules to provide high-energy sensitivity or a triggering
mechanism in high-energy particle experiments.

Avalanche experiments have been carried out on PASS devices with different
geometries, such as regular square configurations immersed in a magnetic field
perpendicular to the plane, or systems of a few lines of microgranules with the
magnetic field applied along the lines.
Results of these experiments show that the simultaneous flip of several granules are
produced in these latter systems with a higher incidence than in the square
configuration \cite{avalanchas}.
The existence of partial avalanches and a lack of transitions of complete lines were
explained as a consequence of imperfections in the manufacturing process
\cite{avalanchas}.

We have performed extensive simulations of transitions in PASS
systems immersed in an increasing external field in both
perpendicular and parallel magnetic field set-ups. Results from
the perpendicular case were presented in Ref. \cite{penaranda5},
showing a reduction of the spread in transition fields, analogous
to
 that experimentally observed.
In the same work, and in agreement with other authors
\cite{Esteve}, the existence of a gap or interval of external
field values for which no transitions occur was shown for
concentrated systems. This corresponds to a hot border, in which
only transitions by increasing temperature can be produced
\cite{girard1}. This gap, which appears in concentrated
configurations, but not in dilute systems, was related to
transitions occurring in preferential positions, which induced
spatial ordering. The presence of this ordering was found to be
consistent with the value of the fraction $f$ of remaining
superconducting microgranules for which the gap appears. This
effect could have relevant consequences on the minimum energy
detected by experimental detectors.

In this work we present the results corresponding to the other case: transitions
in planar arrays with the magnetic field applied parallel to the system.
A gap or plateau zone in the transition field values
will be shown to exist for concentrated configurations, but not for dilute ones,
for microgranules assumed to be microspheres with defects on
their surface. This effect is
similar to that produced in the perpendicular external field case,
but appears for a higher fraction $f$ of remaining superconducting
microgranules.
The avalanche effect is hardly observed in these concentrated systems.
An additional analysis is performed on PASS with more perfect
grains in order to reduce the influence of the defects on the response of
the system. In this case all configurations, both concentrated and dilute, present
plateau zones, but are shown to correspond to two types of different behaviour
 depending on lattice distance. For separated spheres,
transitions are produced
in such a way that remaining superconducting spheres are ordered on
complete lines when the plateau appears.
After this zone, avalanches corresponding to transitions
of these complete lines are observed.
On the contrary, transitions in configurations with smaller lattice
distances produce a quasi-homogeneous spatial
distribution of remaining superconducting spheres in the plateau zone.
Furthermore, there are a smaller number
of avalanches in the evolution of these concentrated systems,
but which involve a larger number of spheres.
We will explain the appearance of the plateau zone and develop a
criterion to determine the lattice distance value that separates the two types of
behaviour.

These results are presented in this paper as follows:
in the next section
transitions undergone by systems corresponding to different lattice
distances are analyzed. Section III  is devoted to studying the behaviour
of more perfect granules.  Spatial order produced by transitions,
the existence of a plateau and the value of the fraction $f$ of spheres
that remain superconducting for which it appears
are explained in this section. The avalanche effect is described in Section IV.
Section V provides our conclusions.

\section{Transitions}

We perform simulations of transitions on systems
of superheated superconducting microgranules in an increasing external
magnetic field $B_{ext}$. Microgranules are considered as
spheres of equal radius $a$, much larger than the London penetration
length.
Results presented in this paper correspond to planar configurations of
spheres ordered on square lattices immersed in an external field
applied parallel to the lines of spheres.
To take into account the existence of possible defects on spheres we
assign a random threshold magnetic field value $B_{th}$ to each sphere
following a distribution related to the superheated value $B_{SH}$.
This distribution takes the form of a parabolic distribution in an
interval of values between $B_{SH}(1-\Delta)$ and $B_{SH}$.
A larger value of delta is
related to a larger range of defects. Results presented in this
section correspond to a $\Delta= 0.2$ value in agreement with
experimental estimations for tin spheres \cite{geigenmuller}.
We then consider that the transition of a sphere occurs when the maximum
surface field at any point on its surface reaches its threshold value,
and that the transition is complete.
We further assume that the latent heat corresponding to transitions is
small enough so as not to affect bath temperature and not to originate
transitions in neighbouring spheres.

The method of calculus in our simulations is as follows: $N$ spheres are
placed at the vertices of a planar square lattice with the lattice
parameter $d$ giving the distance between neighbours.
Without loss of generality, we place the system in the plane $XZ$ and
$(l_{x},l_{z})$ are the coordinates of the spheres,
After assigning  a random threshold field value
to each sphere, the system is immersed in an external magnetic
field $B_{ext}$, applied in the $Z$ direction, which is slowly increased
from zero.
Resolution of the Laplace equation for the magnetic scalar potential
with the appropriate boundary conditions enables the
surface magnetic field to be known on each microsphere.
We have used a numerical procedure that allows us both
to consider the complete multi-body problem and to reach multipolar
contributions of an arbitrary order \cite{Penaranda1}.
When the maximum local magnetic field on the
surface of any sphere reaches its threshold value $B_{th}$,
the sphere transits and
the configuration becomes one of $N-1$ superconducting spheres.
Interactions between remaining superconducting microgranules change
after each transition, due to the long range of diamagnetic interactions.
This leads us to repeat the process until all spheres have transited.
The applied field for which each sphere transits, and the maximum
surface
magnetic field value of each sphere are monitored after each
transition, allowing us to study the evolution of the system in the
successive
transitions.

Numerical simulations have been performed on several
configurations with distances between sphere centres, in units of
radius $a$, of $d/a= 4.376$, $3.473$, $3.034$, $2.757$ and $ 2.5$.
In a 3-D array these distances correspond to values of filling
factors (fraction of volume occupied by the spheres), of $\rho=
0.05$, $0.10$, $0.15$, $0.20$ and $0.268$. The number of spheres
was $N=169$, that is a $13 \times 13 $ square lattice.

\begin{figure}
\includegraphics[width=1.0\textwidth]{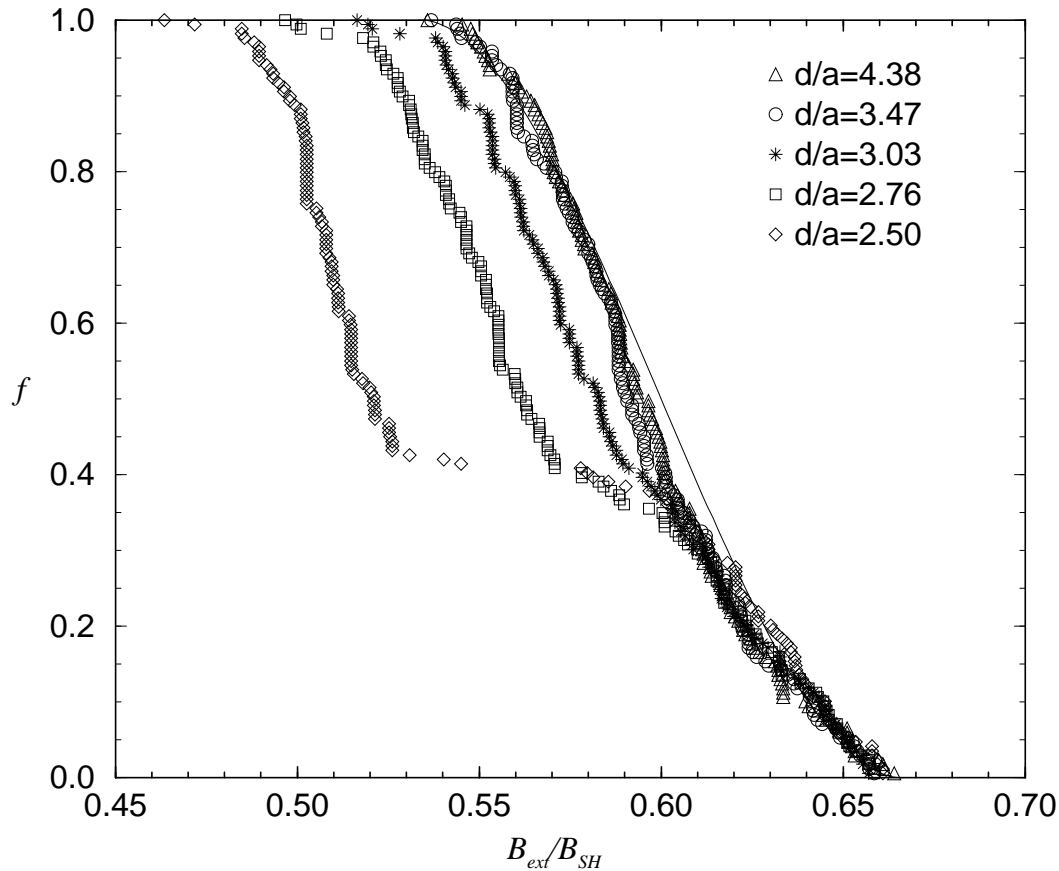}
\caption{Fraction $f$ of spheres that remain superconducting versus
$B_{ext}/B_{SH}$ after an increase of the external field from zero
for several samples of $N=169$ initially superconducting  spheres corresponding
to different lattice spacings. The continuous line corresponds to the dilute limit.}
\label{trans02}
\end{figure}

\begin{figure}
\includegraphics[width=1.0\textwidth]{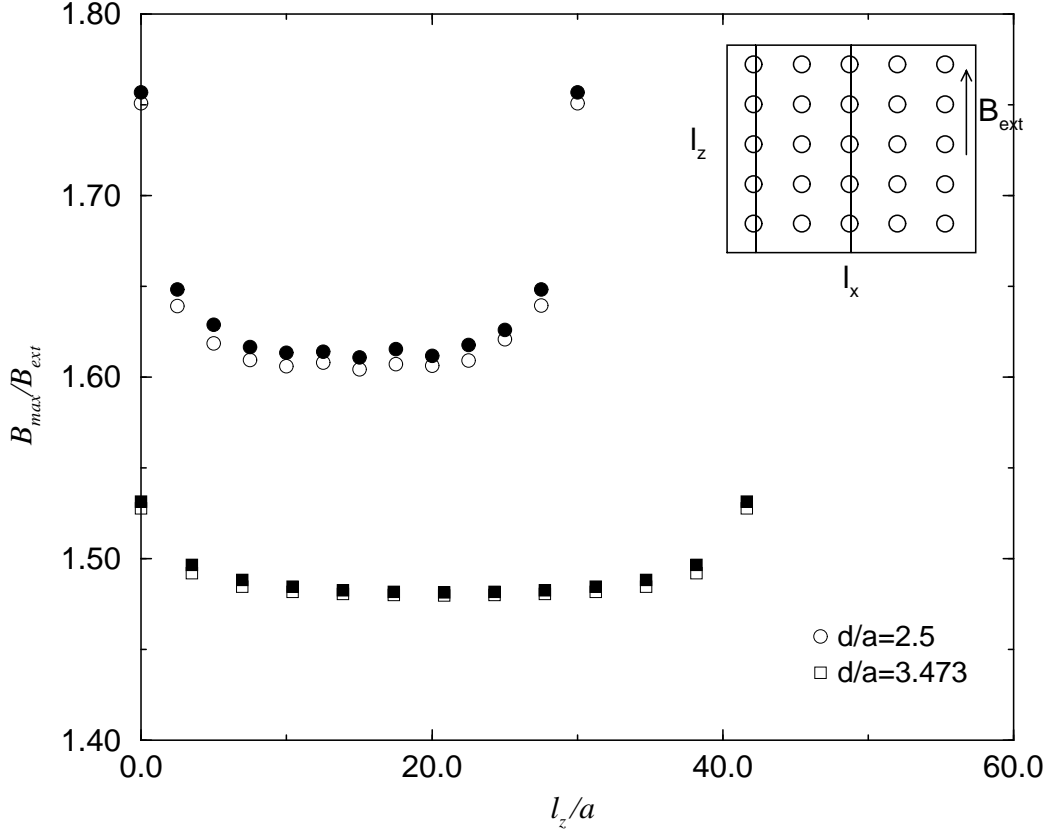}
\caption{Maximum surface field values for spheres on lines parallel to the external
field, as shown in the inset, as a function of the $l_{z}$ coordinate.
 Results correspond to a central line (full symbols)
and a lateral border (hollow symbols).
}
\label{Bmax}
\end{figure}

Results of transition simulations are shown in  Fig. \ref{trans02}.
In this figure the fraction of remaining superconducting spheres
versus the increasing external field (in units of $B_{SH}$) is
represented for several lattice distances. We can observe an
increase in the spread of transition fields as lattice distance
decreases, as a consequence of the stronger diamagnetic
interactions. The  first transitions, produced at the smallest
external field values, correspond in all cases to spheres placed
at the centre of both the top and bottom borders of the system,
which have the maximum surface magnetic field. This is because
diamagnetic interactions between microgranules produce a screening
effect, especially in the direction of the applied field. This can
be seen in Fig. \ref{Bmax}, where maximum surface fields for
spheres on a central line of granules, parallel to the external
magnetic field, and on a line at the lateral border are
represented for two different lattice distances.
However, the most relevant feature of Fig. \ref{trans02} is
the appearance, in concentrated configurations, of a gap or plateau zone
corresponding  to an interval of
external field values for which no transition occurs. This plateau appears
at a fraction of spheres that remain superconducting at approximately $f=0.41$.
In this zone, flips to the normal state can only be induced by thermal
nucleation.  In this sense, this effect corresponds to a hot border
\cite{girard1}.
This behaviour is apparently similar to that observed in systems immersed
in a perpendicular applied magnetic field \cite{penaranda5}, but here the
plateau appears for a higher $f$ value.

\begin{figure}
\includegraphics[width=1.0\textwidth]{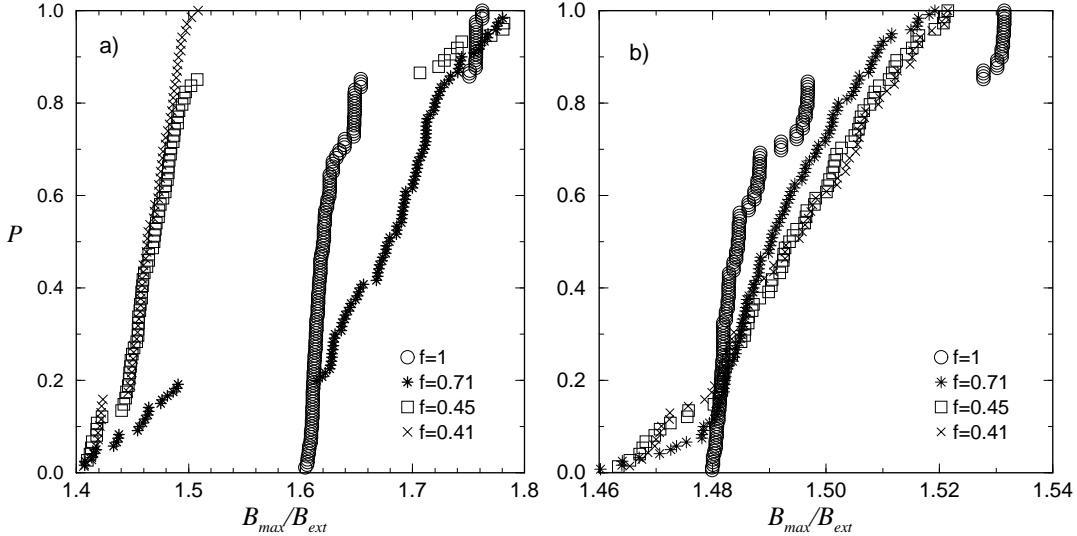}
\caption{Fraction $P$ of spheres with maximum surface field value lower than
the x-value (in units of $B_{ext}$) in the evolution of  configurations with
initial
lattice spacings a) $d/a=2.5$ and b) $d/a=4.376$. Results for $f=1$ (all spheres
superconducting), $f=0.71$, $f=0.45$ and $f=0.41$  ($120$, $74$ and
$69$ spheres that remain superconducting respectively) are shown.}
\label{pbmax02}
\end{figure}

The reason why the plateau appears in concentrated systems, but
not in dilute ones, can be related to the different dynamics in the
evolution of the system during the  transitions.
We have analyzed the dynamics by studying the evolution of both maximum
surface magnetic fields and spatial distributions.
Fig.  \ref{pbmax02} shows such an evolution for two representative cases.
The first, Fig. \ref{pbmax02}.a, is a concentrated configuration
with lattice distance $d/a=2.5$ (that would correspond to a
filling factor of $\rho=26.8\%$ in 3-D systems) and whose
evolution presents the plateau zone. The second, in Fig.
\ref{pbmax02}.b, is a dilute system ($d/a=4.38$, $\rho=5 \%$).
Distributions of maximum surface magnetic field are represented
for both configurations and correspond to values of fraction of
spheres that remain superconducting  $f=1$ (initial state, all
spheres superconducting), $f=0.71$ (120 superconducting spheres),
$f=0.45$, ($74$ superconducting spheres, which correspond to some
transitions before the plateau) and $f=0.41$ (69 superconducting
spheres, when the plateau appears in the concentrated case). It
can be seen that initial distributions present different intervals
of maximum surface magnetic field values that are larger for
concentrated systems and which are a sign of the relevance of
diamagnetic interactions. However, a similar distribution in the
two separated branches is shown for both concentrations. The
smaller branch, with higher surface magnetic values, corresponds
to spheres placed at both the top and bottom borders of the
system. Some of these will be the next spheres to transit.
Transitions change the maximum surface magnetic field values in a
large number of spheres. There is competition between the
increment of surface magnetic field, due to the reduction of the
screening effect, and the decreasing number of interacting
microgranules. For concentrated systems a change in the population
of the two branches is produced, with  an increment in the
distance between both branches. The complete disappearance of the
upper branch causes the system to enter the plateau zone and only
granules with low surface magnetic field values remain
superconducting, and are thus far from their threshold values.
They need much larger external fields to flip to normal state.
This  explains the appearance of the plateau. On the contrary, the
two initial branches merge and change to a more continuous
distribution after some transitions in dilute systems. Slight
increments of the applied field are able to produce subsequent
transitions and therefore no gap interval is observed.

\begin{figure}
\includegraphics[width=1.0\textwidth]{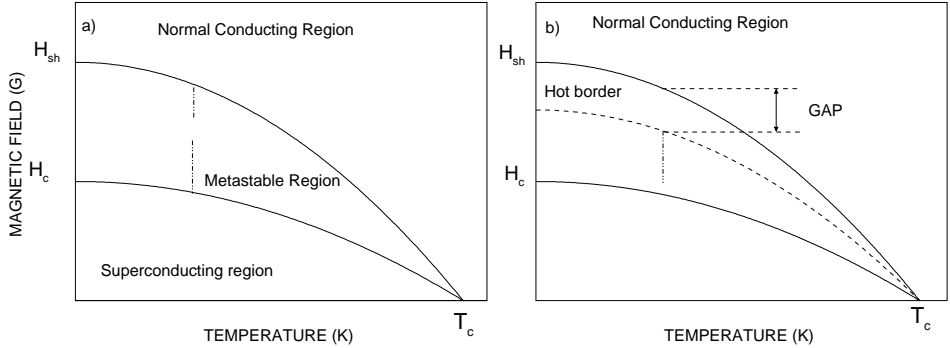}
\caption{a) Phase diagram corresponding to initial PASS configurations. b) Phase
diagram of remaining superconducting spheres, in concentrated systems, when the
plateau appears.}
\label{diagrama}
\end{figure}

Relating these types of behaviour with the position of the ensemble of spheres
in a phase diagram, this evolution could be described in the
following way: the initial state could be represented by a set of points
along a vertical line at the bath temperature,
distributed into two separated segments, one of which is near the superheated
line (Fig. \ref{diagrama}.a).
Each point  represents the state (T,H) of a sphere  and,
consequently, the upper segment shows the population
corresponding to the upper field branch. The separation between the segments
coincides with the separation between both branches.
Changes in diamagnetic interactions
by transitions produce a change in the population of each segment.
In dilute configurations a connection of both
segments is produced and the phase diagram presents a population with an
approximately continuous distribution below the superheated line.
On the the other hand, transitions in concentrated systems are produced in
such a way that the two segments remain
separated until the upper segment disappears (Fig. \ref{diagrama}.b).
It is at this stage that the plateau zone appears. Remaining
superconducting spheres correspond to the population
of the remaining segment, separated from the superheated line. This effect, for
each
temperature, describes a hot border line in the phase diagram separated from
the superheated line by an interval of magnetic fields corresponding to
the plateau zone gap.

\begin{figure}
\includegraphics[width=1.0\textwidth]{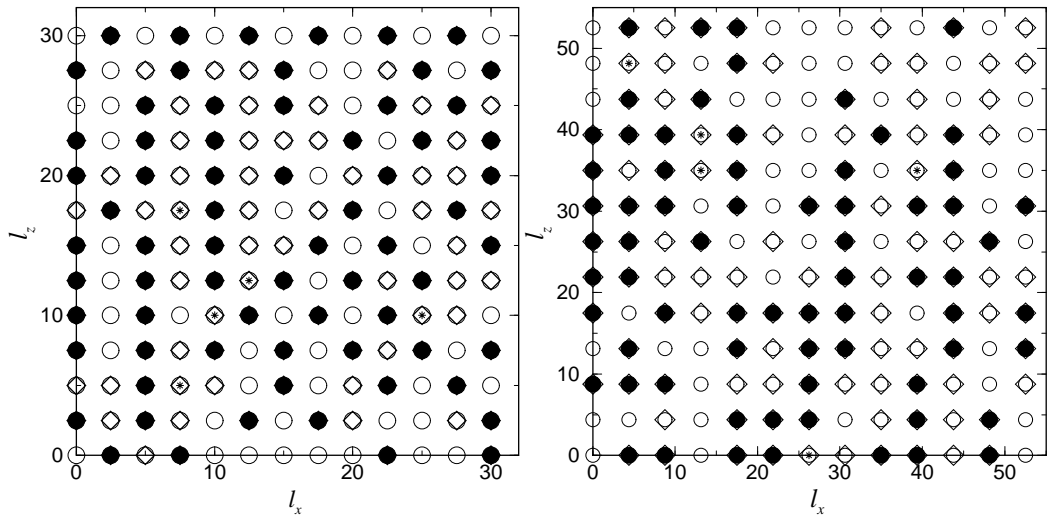}
\caption{Spatial distributions of superconducting spheres in
 $13 \times 13 $ systems with lattice distances
$d/a=2.5$ (left) and $d/a=4.376$ (right),
for $f=0.71$ (diamonds), $f=0.45$ (diamonds with dots) and $f=0.41$ (full symbols)
($120$, $74$ and $69$ remaining superconducting spheres respectively).
}
\label{pos02}
\end{figure}

We have studied the relation between surface magnetic distributions
and spatial distributions of the remaining superconducting spheres.
The evolution of these spatial distributions present distinct features
depending on the concentration
of the system, as is shown in  Fig. \ref{pos02} for  two representative cases
($d/a=2.5$ and 4.376 respectively).
It can be observed that transitions in concentrated systems produce,
when the plateau appears, two separated domains with different kinds of
spatial order.
For both domains the tendency is not to have any superconducting
next neighbour in the direction perpendicular
to the external field. While spheres of the first domain are mainly
placed along vertical lines,
a tendency to align along the diagonal direction is observed for the other domain.
Dilute systems only show partial lines of superconducting
spheres.

\section{ Defect-free granules}

Results presented above correspond to spheres with a relatively large
range of defects. It is well known that defects can act as
nucleation centres and therefore transitions are
produced at smaller external fields than in more perfect spheres.
This fact could have an influence on the dynamics of the system,
because it can alter
spatial distributions and thus diamagnetic interactions between
spheres after transitions.

In order to gain insight into the intrinsic dynamics of the
evolution of these systems, we have reduced the influence of
defects and analyzed the response of systems formed by nearly
perfect spheres. Results are obtained following the same calculus
protocol, but the threshold magnetic field distribution is changed
in a smaller interval, that is using a smaller $\Delta$ value. In
this section the value selected is $\Delta=10^{-6}$.

\begin{figure}
\includegraphics[width=1.0\textwidth]{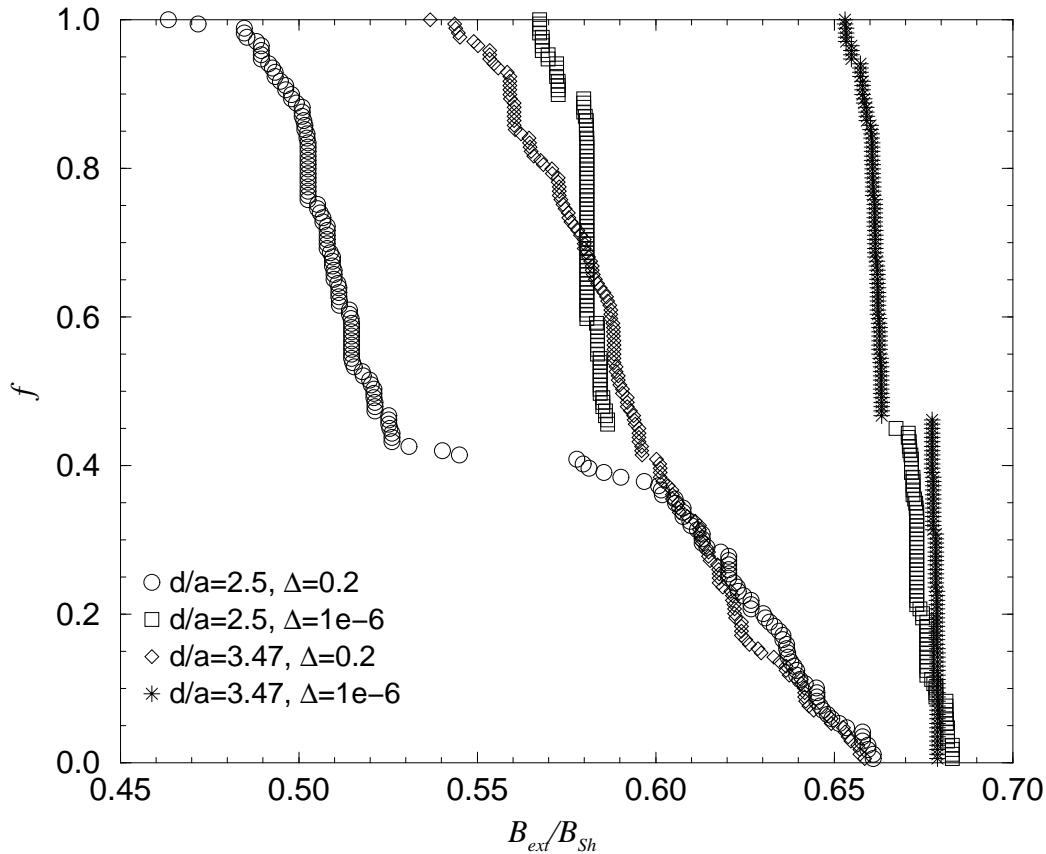}
\caption{Fraction $f$ of remaining superconducting spheres versus
$B_{ext}/B_{SH}$ after an increase of the external field from zero,
for several samples of $N=169$ initially superconducting  spheres corresponding
to different lattice spacings and defect distributions. Results obtained for
$\Delta=0.2$ and
$10^{-6}$ are represented for lattice spacings $d/a=2.5$ and $3.473$.
}
\label{difdelta}
\end{figure}

Fig. \ref{difdelta} displays the fraction $f$ of remaining superconducting
spheres as a function of the external field for two lattice distances,
$d/a=2.5$ and $3.473$ ($26,8\%$ and $10\%$ respectively),
comparing the results obtained for $\Delta=0.2$ and $10^{-6}$.
The reduction of the spread
in transition fields for the lower delta value is observed in this figure,
although the influence of
diamagnetic interactions is clearly seen in the wider interval of transition
fields shown by the  more concentrated systems.
On the other hand, the more discontinuous aspect of these transition lines for
$\Delta=10^{-6}$ suggests the possibility of a more important avalanche effect.
A further significant feature
in the figure is the appearance of a plateau zone for dilute
configurations.
Calculations on different lattice distance values show the existence
of such plateau zones in all configurations. This zone is larger for small
lattice distances. In addition, the plateau is produced near the $f=0.45$
value, {\it i.e.} slightly higher than in spheres with defects.
This plateau appears in dilute systems for a
fraction of remaining superconducting spheres of about $f=0.47$.

\begin{figure}
\includegraphics[width=1.0\textwidth]{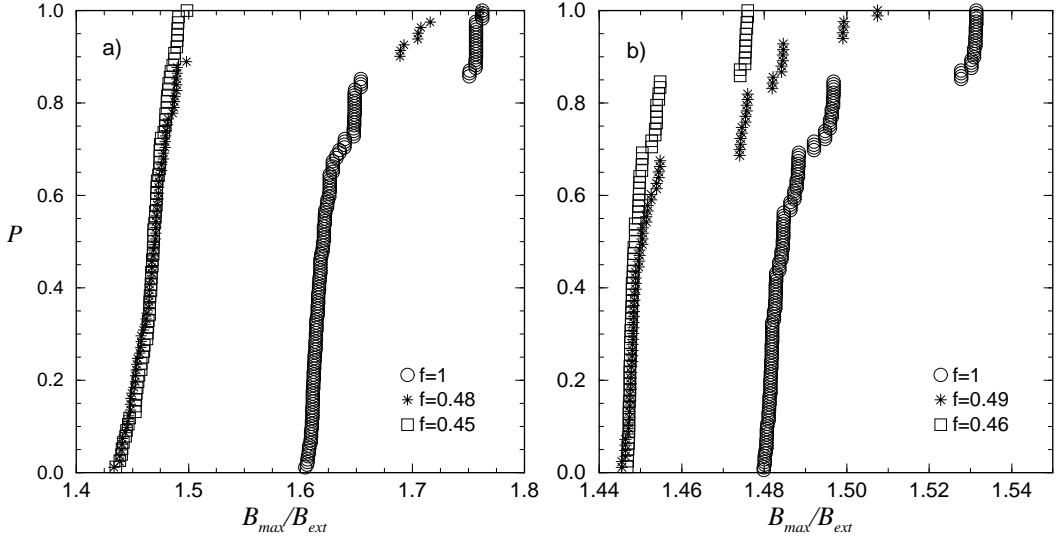}
\caption{Fraction $P$ of spheres with maximum surface field value lower than
the x-value (in units of $B_{ext}$) in the evolution of  configurations with
initial lattice spacings a) $d/a=2.5$ and b) $d/a=3.473$.
Results correspond to the initial distribution ($f=1$) and those corresponding to
several transitions before the plateau ($f=0.48$) and at the plateau zone
($f=0.45$).
}
\label{bmx10-6}
\end{figure}

This change in the behaviour of the system due to the influence of defects,
especially in larger lattice distances, and the value of $f$ for the
appearance of the plateau, can be related to
 the different dynamics shown in their evolution,
reflected in both spatial and surface magnetic field distributions.
Maximum surface field distributions corresponding
to systems with lattice distances $d/a=2.5$ and $3.473$ are displayed
in Fig. \ref{bmx10-6}.
The concentrated case (Fig. \ref{bmx10-6}.a)
presents an evolution similar to those of spheres with defects,
showing the two branches until the disappearance of that corresponding
to higher surface field values when the plateau appears.
The only differences are the slightly smaller interval of values of
the surface magnetic field and the fact that the upper branch disappears
at a larger value of $f$.
The dilute configuration (Fig. \ref{bmx10-6}.b) presents, on the other hand,
a qualitative change in its evolution. The two branches of the initial
distribution split into several close branches after successive transitions.
In the plateau zone ($f=0.46$) the surface magnetic fields of the remaining
superconducting spheres are in a smaller interval, quite far from their threshold
values. A larger increment of the applied field is therefore necessary to produce
the next transition. This is the reason for
the appearance of the plateau in this case. This gap in external field values
originates a hot border zone analogous to the concentrated case. Only
a temperature increase in this zone can originate a flip to the normal state.

\begin{figure}
\includegraphics[width=1.0\textwidth]{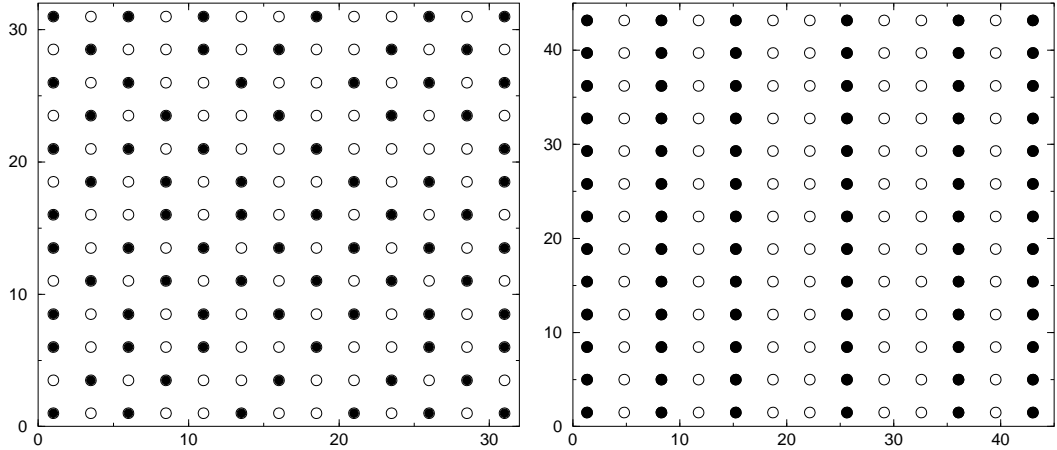}
\caption{Spatial distributions of initial $N=169$ spheres with lattice distances
$d/a=2.5$ (on the left) and $d/a=3.473$ (on the right), and those corresponding
to the plateau zone.
}
\label{pos10-6}
\end{figure}

The changes in magnetic surface field distributions, especially
for dilute systems, suggest a change in the evolution of spatial
distributions with respect to less perfect spheres. The study of
spatial distributions during transitions reveals interesting
behaviour. The more perfect spheres present ordered configurations
at the plateau, both for concentrated and dilute systems, but the
order is clearly different, as is shown in Fig. \ref{pos10-6}.
Evolution in dilute systems is produced
in such a way that, when the plateau appears, remaining superconducting
spheres display a spatial disposition in complete lines parallel to the external
field alternating with complete lines of spheres in the normal conducting
state. This spatial distribution explains the $f$ value for which the plateau
appears. Perfect alternation of superconducting and normal lines would
correspond to a $f=0,5$ value. A small deviation of perfect alternation is
coherent with the $f=0.47$ value obtained for these dilute systems.
Repeating simulations for different
lattice distance values, we observe that this kind of order is achieved
for configurations with lattice distance values $d/a=3.034$ and larger,
but not for $d/a=2.757$ or $2.5$.

For more concentrated systems
the remaining superconducting spheres appear on the plateau
aligned in the diagonal direction, $\it {i.e.}$ with next neighbours
in the normal state, but with superconducting second neighbours.
Hence, microgranules present a square lattice configuration with a
parameter of $\sqrt{2} d$ value.
This fact corresponds to a $f=0.5$ value for the plateau zone.
Defects in this ordering can be seen as boundaries between perfectly
ordered domains and their effect is to slightly reduce this value to the
obtained $f=0.46$.

\begin{figure}
\includegraphics[width=1.0\textwidth]{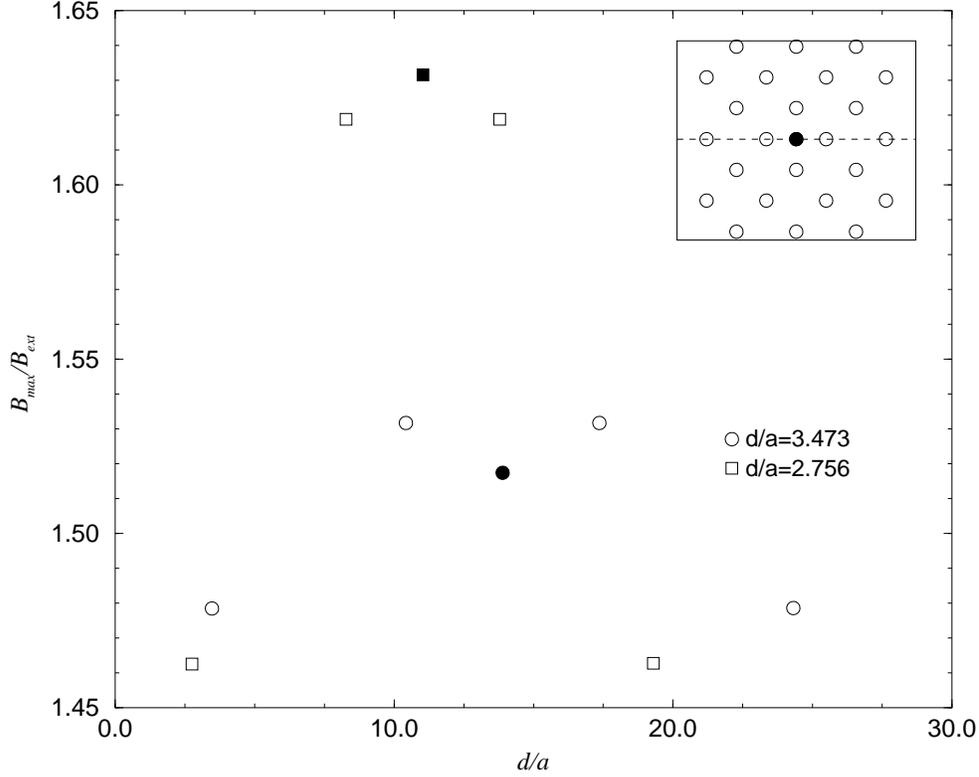}
\caption{Maximum surface magnetic field values (in units of $B_{ext})$
for configurations prepared as shown
in the inset. Field values on spheres lying on
a line containing the central sphere are represented for  $d/a=2.5$ and
$d/a=3.473$.
}
\label{crite1}
\end{figure}

\begin{figure}
\includegraphics[width=1.0\textwidth]{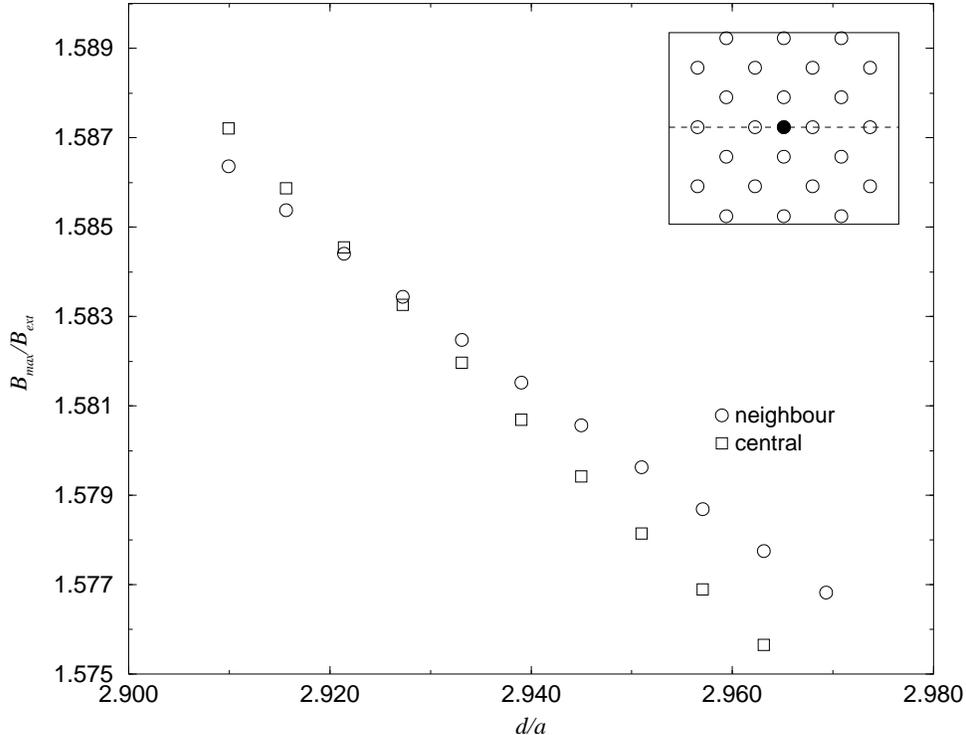}
\caption{Maximum surface magnetic field values (in units of $B_{ext}$)
versus lattice distances, corresponding to the central sphere and its nearest
horizontal neighbour in the prepared configuration.
}
\label{crite2}
\end{figure}

The existence of one or another kind of order depends on the
concentration of the systems.
We have developed a criterion to determine the limit of lattice
distances that
separates the two different dynamics. For this criterion we consider a
configuration similar to that of the domains present in concentrated systems.
We have simulated systems of $81$ superconducting spheres placed on a  
$9 \times 18$ array in the diagonal configuration.
An additional superconducting sphere is placed in the middle of the 9th row, 
{\it i.e.} nearly in the centre of the system and with 4 nearest neighbours.
A scheme of this configuration is presented in the inset of Fig.
\ref{crite1}.
The extra grain should be the first to transit to achieve
the order present in concentrated systems. In another case this order
is not possible.
We have calculated the maximum surface magnetic field values on
all  spheres of this
configuration for several values of lattice distances. Results of these
simulations are shown in the
Fig   \ref{crite1}  for  domains corresponding to
$d/a=3.473$ and $2.756$ respectively. In this figure maximum
surface field values on the extra sphere (full symbols)
and the values corresponding to spheres with the same $l_{z}$ coordinate,
shown by the dashed line in the inset
of the figure, are represented.
It can be observed that, for the lower lattice distance value,
the extra sphere has a higher surface magnetic field than
its neighbours so it will be the next one to transit. Consequently, the
configuration
can present the order observed in concentrated systems. On the other hand,
for $d/a=3.437$ the maximum surface magnetic field on the central sphere is
smaller
than on its neighbours, one of which will be the next to transit, so it is thus
impossible
to achieve such an ordering of the domain.
By repeating simulations for additional intermediate lattice distances we can find
the limiting value for which the change of behaviour is produced. We obtain this
limit for a distance between sphere centres of
$d/a=2.921$ ($\rho=16.8\%$). Results displayed in Fig. \ref{crite2} near this
limit clearly show this change.

\section{Avalanche effect}

Avalanches can occur in SSG systems.
The transition of any microgranule
changes diamagnetic interactions in the system, especially between its neighbouring
granules. Depending
on the relative position of the microgranules and on the direction of the
applied field, a transition of a single grain can then produce the flip of other
microspheres to the normal state.

This effect can be useful in  detecting low-energy radiation.
To produce the transition of a microgranule to the normal
state  a quantity of energy proportional to the volume of the
grain is necessary in the global heating model \cite{Dubos}.
In this sense, a device composed of small microspheres could act
as an ultrasensitive low-energy detector. The change of magnetic flux
induced by the loss of the Meissner effect could be
amplified if an avalanche of transitions is produced. The enhancement of the
signal
would permit the detection of the incident particle or radiation.
In addition, the amplified signal could be used as a trigger in high-
energy experiments.

PASS systems with the external field parallel to the plane seem suitable
systems to show avalanches.
Diamagnetic interactions can favour the screening
effect of a granule on neighbours, mainly on those placed on lines parallel to
the external field. The maximum surface magnetic field is, in consequence, lower
on internal spheres than on the spheres at the top and bottom borders.
Transition of a sphere to the normal state allows the introduction of magnetic
field lines around their neighbours due to the reduction of screening, and increases
their surface magnetic field. Some of the spheres can achieve their
threshold magnetic field values and transit simultaneously producing an
avalanche.

\begin{figure}
\includegraphics[width=1.0\textwidth]{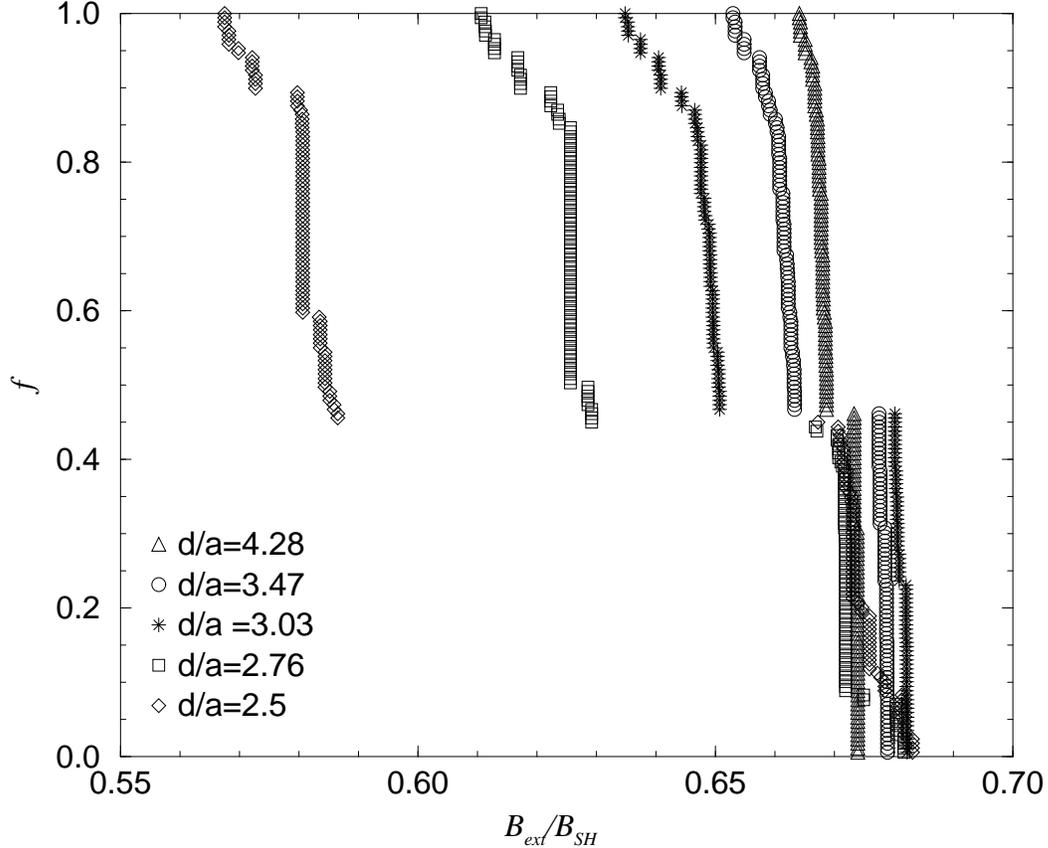}
\caption{Fraction $f$ of spheres that remain superconducting versus
$B_{ext}/B_{SH}$ after an increase of the external field from zero
for several samples of $N=169$ initially superconducting  spheres corresponding
to different lattice spacing. Results correspond to defect-free spheres.
}
\label{f1-6}
\end{figure}

Results from simulations on systems with defect-free spheres are coherent
with the existence of avalanches.
Fig. \ref{f1-6} shows the fraction of remaining superconducting spheres
in an increasing external field for several lattice distance
values.
The shape of transition lines indicates that a great number of
transitions are produced at very close values of the applied field
indicating an avalanche effect.

A deeper study shows that first avalanches correspond, in all configurations,
to simultaneous transitions of two or four spheres  placed
at the borders, reflecting the top-bottom and left-right symmetry.
After this initial transition, systems present clearly
differentiated behaviour depending on the lattice distances.

In dilute systems the next avalanches, of $2$ or $4$ spheres, are only produced
in  lines that have undergone previous transitions. They include transitions
of both top- and bottom-ending superconducting spheres.
When $3-4$ spheres at both ends have transited on each of these lines,
an avalanche of the remaining superconducting spheres of the line takes place.
This situation corresponds to the appearance of the plateau zone, with a spatial
distribution of only complete lines of  spheres that remain superconducting.
An increment of the external field larger than the gap value generates successive
avalanches of complete lines.

In concentrated systems, on the other hand, initial avalanches of spheres at the
borders
are followed by a small number of simultaneous transitions of $2$ or $4$ spheres
in alternate sites near the top and bottom borders.  A further
small  increment of the applied field produces a large avalanche. This avalanche
includes a large number of spheres following the same alternate pattern.
After this avalanche the plateau zone appears.
Remaining superconducting spheres are located in several domains
with regular square distributions.
The next  avalanches correspond to  simultaneous transitions of spheres
from each domain.

\begin{figure}
\includegraphics[width=1.0\textwidth]{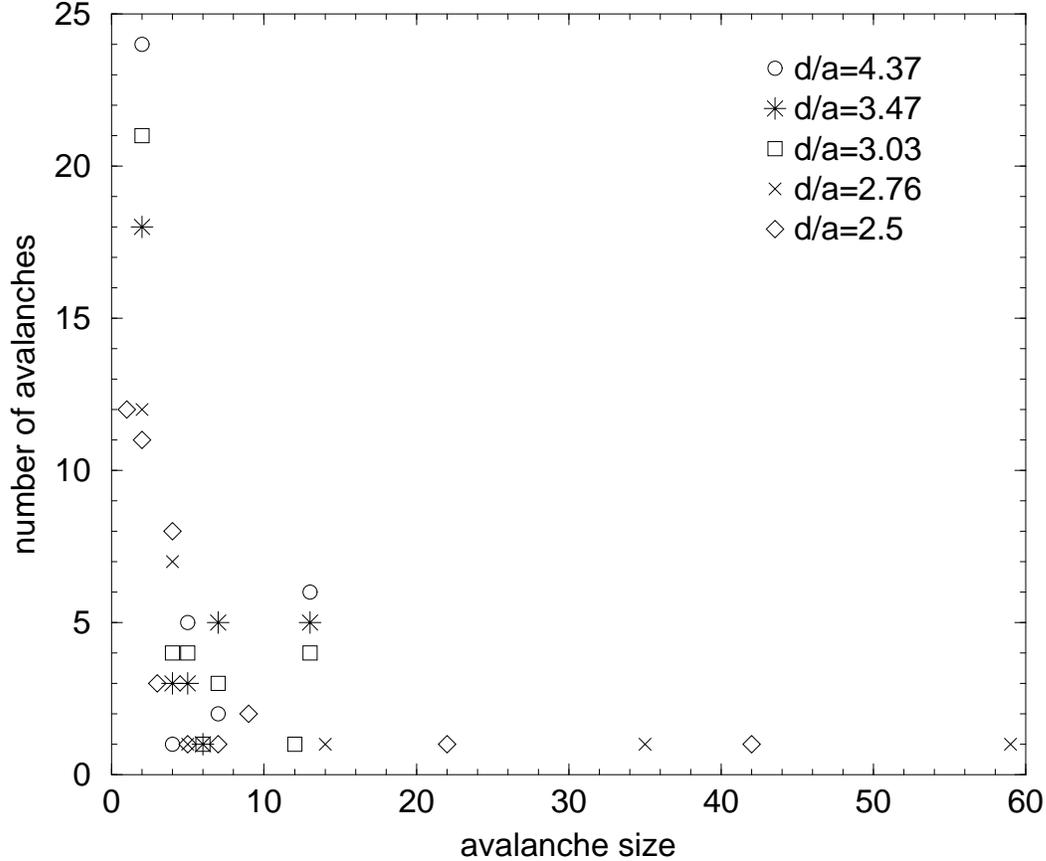}
\caption{Number of avalanches of defect-free spheres versus avalanche size
for several lattice distances.
}
\label{avalancha}
\end{figure}

In Fig. \ref{avalancha} the number of observed avalanches versus their size
(the number of spheres that are involved) are represented for several lattice
distances.
Results show that large avalanches only appear for concentrated systems. On the
contrary, dilute configurations show a higher number of small-size avalanches.

Experimental devices can make use of this effect by placing the detector
in the desired zone of transition lines.
Knowledge of the different types of behaviour of the system enables selection of
the type of configuration according to the required size of avalanche.
It should be pointed out that the experimental precision could not be sufficient
to resolve very close avalanches, which in consequence would be detected as
a single flip.

Experimental results presented by Meagher et al \cite{avalanchas} correspond to
a system of separated lines ($l/a=16.66$) of closed spheres ($d/a=2.22$).
It is thus a type of configuration different from that studied here and could be
considered to be a
hybrid of both the concentrated and diluted configurations. It could therefore
show intermediate behaviour.
This could explain the absence of avalanches of complete lines observed in
these experiments.

This study can be extended to other geometrical configurations of
superheated  superconducting granules than can favour some types of
multiple flip. Simulations of experimental systems and different superpositions
of planar structures are presently in progress.

\section{Conclusions}

Simulations of transitions in PASS systems immersed in a parallel external field
have been performed for several lattice distance values.
Spheres with defects on their surface, which can act as nucleation centres,
and quasi defect-free spheres are considered.
A plateau zone is observed in concentrated systems for both kinds of spheres.
This zone corresponds to a gap in the external
field values that are able to produce transitions. On the other hand, for large lattice
distances the plateau only appears for more perfect spheres.
This plateau corresponds to a hot border, in which only transitions by a finite
temperature increment can be produced. This effect can have relevant
consequences in experimental detectors due to the fact that  the uncertainty in the energy
threshold for transitions can be reduced. On the other hand, transitions in this
zone could characterize
the emission spectrum of a source of radiation.

The existence of the plateau
and the fraction of remaining superconducting spheres for which it appears
are explained as a consequence of two different dynamics
during transitions, for dilute and concentrated configurations respectively,
which lead to two different spatial orders.
Systems with small lattice distances present, at the plateau zone,
a distribution on domains where the superconducting
spheres are homogeneously located.
Dynamics for larger lattice distances generates, at the plateau zone, an ordering
of spheres that remain superconducting on complete lines parallel to the
external field at the plateau zone. This ordering is consistent with the
fraction value of
remaining superconducting spheres for which the plateau appears.
A criterion has been developed in order to find the limit between the two
distinguishable dynamics. This limit is obtained for a distance between
spheres of $d/a=2.921$ ($\rho=16.8\%$ in 3-D systems).

A study of the avalanches produced in these systems shows distinct types of behaviour
depending on the lattice distance values. Concentrated systems present, after
small initial avalanches corresponding to the first transitions, a large avalanche
before
the gap is reached. After this zone, successive avalanches corresponding to
transitions of spheres from each domain are observed.
Dilute configurations show relatively small avalanches before the plateau appears.
After this zone, successive avalanches of complete  lines are produced.
These effects could be used in experimental devices to detect low-energy 
radiation
that can be amplified by  multiple transitions.

\section*{Acknowledgments}
We would like to thank T. Girard for helpful discussions.
We acknowledge financial support from Direcci\'on General de
Investigaci\'on Cient\'{\i}fica y T\'ecnica (Spain)
(projects BFM2003-07850-C03-02 and BFM2002-02629) and
Comissionat per a Universitats I Recerca (Spain)
(project 2001SGR97-0021).
We also acknowledge computing support from Fundaci\'o Catalana per a la
Recerca-Centre de Supercomputaci\'o de Catalunya (Spain).

\end{document}